\documentstyle[12pt]{article}
\newcommand{\journal}[4]{{{\sl #1}} {\bf #2}, {#3} (#4)}
\newcommand{\prb}[3]{\journal{Phys.~Rev.~B}{#1}{#2}{#3}}
\newcommand{\prl}[3]{\journal{Phys.~Rev.~Lett.~}{#1}{#2}{#3}}

\newcommand{\cutau}{Cu$_{\rm 3}$Au}

\def\C44{C_{\rm 44}}
\begin{document}
\draft
\begin{title}
Kinetics of Ordering\\
in \cutau\ : An Atomistic Approach
\end{title}
\author{Zhigang Xi and Bulbul Chakraborty}
\begin{instit}
Physics Department\\
Brandeis University\\
Waltham, MA 02254, USA
\end{instit}
\begin{abstract}

We study the kinetics of ordering in \cutau\
using a model Hamiltonian derived from the effective medium theory of
chemical bonding.
Monte Carlo
simulations are used to investigate universal and non-universal features of the
growth kinetics.  Anisotropic scaling of the structure factor is
observed in late-stage growth of ordered
domains.  The anisotropy is a non-universal feature determined by the
details of the microscopic model, and we find that the anisotropy
observed in the simulations is
in excellent agreement with experiments on \cutau.  The simulations are
discussed
in the context of theories of unstable growth.
To our knowledge, this is the first study of kinetics in
a realistic model Hamiltonian describing the material-specific
properties of \cutau.
\end{abstract}
\draft
\section{INTRODUCTION}
Kinetics of ordering is a fundamental problem in the statistical mechanics of
nonlinear phenomena occurring far from equilibrium.\cite{Gunton}
Binary alloys, such as
\cutau, when quenched from a high temperature disordered state to a temperature
below the order-disorder transition temperature, evolve from an initially
metastable or unstable state\cite{Gunbook} towards the stable ordered
state. During the early
stages of growth, the free energy difference between the disordered and ordered
phases and the nucleation barriers between these phases determine the
transformation into ordered regions.  This stage of the growth process is
expected
to be system specific.   In contrast, the late stages of growth, where the
system
consists of ordered domains separated by domain walls, are expected to be have
a
more universal character and there have been various indications that there are
universal scaling laws determining this late stage growth.\cite{Gunton,Mazenko}
{}From a practical viewpoint a knowledge of the morphology and growth
of
metastable
and unstable states is crucial to metallurgy.
The theoretical study of growth kinetics in alloys have
been based upon simulations of kinetic Ising models or simple Langevin
dynamics defined
by the Time-dependent Ginzburg-Landau (TDGL) model.\cite{Mazenko}

There are fascinating experimental results in \cutau.  The
experiments probing late stage growth are in agreement with general ideas of
scaling and universality but exhibit large anisotropy of the structure
factor.\cite{Nagler}
Experiments probing the early stages of growth\cite{Ludwig}
can distinguish between nucleated and continuous ordering and are more
sensitive to the characteristics of a specific alloy through features like the
activation barriers to nucleation and the approach to the classical spinodal.
These experiments have
have shown some unusual growth features that are not completely well
understood.\cite{Ludwig}

A question that intrigued us was whether a realistic model of cohesion and
stability in intermetallic alloys such as \cutau\ would lead to a description
of
growth kinetics which were in essential agreement with the ideas of
universality
and the classification of universality classes derived from the simple models
of
binary alloys, and what were the crucial
non-universal characteristics distinguishing one intermetallic alloy from
another.  The model Hamiltonian used in this study was constructed from the
Effective Medium Theory of cohesion in metals (EMT)\cite{dth} and has
been described in
detail elsewhere.\cite{jphys,bcprl,xithesis}  It has been shown that the model
provides an excellent description of the equilibrium statistical mechanics of
Cu-Au alloys.\cite{jphys,bcprl}
The current study extends the application of this model to
phenomena occuring far from equilibrium.  We have concentrated on the
description
of late stage growth with the intent of defining the universality class for
this
model and to verify the applicability of universality and scaling.

The essential features of our simulation of growth kinetics can be
summarised by saying
that there is evidence for anisotropic scaling with the growth law for domain
size obeying a power law close to that expected for ordering
systems.\cite{Gunton}

\section{SIMULATION OF LATE STAGE GROWTH }

Of the simple ordering  alloys, \cutau\ is one of the more interesting ones
even
when described by a kinetic Ising model.  The groundstate of this model is
four-fold degenerate and therefore it does not belong to any of the well
established universality classes\cite{Gunton,Mazenko,Bindersadiq}  TDGL models
based upon symmetry arguments and phenomenology have a richer structure  with
a three component order parameter and anisotropic gradient terms.\cite{Lai}  We
are unaware of any simulations of the kinetic Ising model appropriate to
\cutau,
but the results of our simulations are in essential agreement with the study of
the TDGL model\cite{Lai} and with the late-stage growth
experiments.\cite{Nagler}

We study growth kinetics of \cutau\ by performing Monte Carlo simulations based
upon the EMT Hamiltonian.  Beside the basic Monte Carlo move of swapping the
identity of a pair of atoms (a move that preserves alloy concentration), we
allow
the computation box to breathe and change aspect ratios as was done in our
simulations of equilibrium phenomena.\cite{jphys}
This dynamic model does not allow local
relaxations or vibrations.  Therefore this is a minimal extension of the
kinetics
associated with Ising models.  The interactions in the EMT Hamiltonian depend
upon the local environment of an atom determined by the type of the surrounding
atoms and the distances.
We do not map the EMT Hamiltonian onto an Ising model. The distance dependence
of
the interactions would neccessarily lead to a an Ising model with temperature
dependent interactions if we do carry out such a mapping.\cite{jphys}

The simulations are carried out in a box with linear dimension of 30 times the
lattice parameter along each direction and containing 108,000 atoms.  The alloy
is annealed at an extremely high temperature (10,000K) and then quenched to a
temperature of 540K (the transition temperature is 612K). The developement of
order is monitored by storing the configuration after every 100 Monte Carlo
steps
(10 initial steps are discarded)  A series of snapshots showing the growth of
ordered domains is shown in Fig. ~\ref{fig:snap}

\begin{figure}[tbhp]
\vspace{5.5 in}
\caption{Two dimensional cuts showing the transformation of the unstable
disordered state into regions of the stable ordered phase.  The four different
symbols the four equivalent ordered states.  The time of taking
the snapshot is indicated at the top of each frame.}
\label{fig:snap}
\end{figure}

The initial stages of growth
indicate that the disordered state is unstable to long-wavelength fluctuations
an
that there is no metastability of the disordered phase at this temperature.
The
three frames at 610 MCS (Monte Carlo step/site) show the three faces of the
cube
and together they convey a picture of interconnected domains that are truly
three-dimensional.  The pictures of domains taken at the end of the simulation
show that each ordered domain is very nearly cubic, and the domain walls have
very little curvature.
The stability of cubic domains is consistent with the picture of
curvature-driven
domain growth\cite{Allencahn} where the velocity of the domain walls is
determined by their local curvature.  These snapshots can, however, only give
us a
qualitative idea of the domain growth process and only quantitative statistical
analysis can lead to a description of growth laws existence of scaling.  This
analysis is most conveniently carried out in terms of the structure factor.
This
also bears the closest relationship to experiment.  The growth of order is
associated with the growth of superlattice Bragg peaks.  A natural length scale
to associate with domain growth is the inverse of the width of the Bragg peak.
A different definition of this length scale is provided in terms of the sum of
the scattering intensities at the three superlattice Bragg peaks correponding
to
the three concentration waves or order parameters.\cite{Bindersadiq}  We have
used both to obtain an estimate of the power law associated with domain growth.
As will be shown below, the structure factor is extremely anisotropic and we
have
analyzed three different cuts through the Bragg peaks.  There are fluctuations
from one simulation run to another and averages over many different runs are
needed in order to obtain a realistic statistical description.  All our results
were obtained by averaging over ten different runs.

\section{SCALING AND ANISOTROPY OF STRUCTURE FACTOR}

We present contour plots of the structure factor along the transverse and
radial
planes passing through a Bragg peak in Fig. ~\ref{fig:str1}.  This figure also
shows cuts along two different directions in each of the planes.

\begin{figure}[tbhp]
\vspace{4.5 in}
\caption{Structure factor obtained by fourier transforming one atomic
configuration.  The anisotropy is clearly demonstarted by the contour plots.
The
small k contours in the transverse plane show a square shape but the rapidly
evolve into circular contours and the overall shape of the structure factor is
the disc shape observed in experiments.  The dashed lines are cuts along
(Q,k,0),
where Q denotes the position of the superlattice peak.  The dip at k=0 is
caused
by a numerical approximation that does not affect any other results.}
\label{fig:str1}
\end{figure}

The anisotropy is directly related to the difference in energy between the two
types of domain walls in \cutau.\cite{Lai}  This can be shown from a linear
stability analysis of the appropriate TDGL equations.  We determine
the  ratio of the peak
widths along the radial and transverse directions, $k_c^{R}/k_c^{T}$,  to be
$\sim 2.5$, in excellent agreement with the experimentally measured
value.\cite{Nagler}  This demonstrates that EMT provides a very good
description
of interfacial energies and is an excellent model for studying defects and
associated phenomena.
determined to be

The line shape of the structure-factor cuts are best fitted by a Gaussian.
However, the average radial or transverse structure factor, obtained by
integration over the transverse plane are much closer to Lorentzian square
shapes.  The Lorentzian square shape leads to a scattering intensity decreasing
as $k^{-4}$ (Porod's law).\cite{Lai,Nagler}   The power laws obtained for
domain
growth obtained by monitoring the width of the superlattice peaks is shown in
Fig. ~\ref{fig:str2}.

\begin{figure}[tbhp]
\vspace{4. in}
\caption{The inverse of the width of the Gaussian line shapes (which give a
measure of the domain size) plotted as a
function of time. It is clear that during the late stages of growth, the
inverse
width can be described by a power law. The crosses are for the radial cut and
the
other set of symbols show the two different transverse cuts, (Q,k,0) and
(Q,k,k).}
\label{fig:str2}
\end{figure}

We observe the same power law growth in the transverse and radial directions
and
obtain a growth exponent of $\sim 0.4$.  This is smaller than the
Allen-Cahn exponent of
0.5.\cite{Gunton,Allencahn}  However, without analysing the effect of averaging
over many more runs and the nature of finite-size distortions in this extremely
anisotropic system we cannot say with certainty that the growth exponent is
significantly different.  The anisotropy makes it more difficult to interpret
the
growth laws since the power law behavior is expected to hold only when the
domain
size is much larger than the domain wall width.
The widths of the
domain walls are determined by the interfacial free energy which is different
for
the two types of walls.  We could therefore be in a scaling regime for one type
of domain wall but not for the other and this could affect our power laws.  We
are carrying out detailed investigations to sort out these questions.  Analysis
of the intensity of the sum of the intensity of the Bragg peaks shows a
crossover
from a power of 0.5 to 0.33 .  This could simply be a finite size effect or
could
be an indication of new physics arising from the coupling to the conserved
concentration variable.\cite{Bindersadiq}

A remarkable observation in the field of growth kinetics has been the existence
of a scaling law for the structure factor:  $S(k,t) = (1/k_c)^d F(k/k_c)$.
Here
$k$ is the magnitude of the wave vector and $k_c$ is the inverse length scale
(for
example the width of the Bragg peaks) which is expected to have a power law
behavior, and $F(x)$ is a scaling function.  The physical picture
behind this scaling form
is analogous to static critical phenomena where the correlation length is the
largest length scale and as a consequence all quantities in te critical region
obey scaling with power laws depending on universality classes.  The
universality classes for growth kinetics are still not well established nor is
the existence of scaling.  We have investigated scaling in our model of \cutau\
and the results are shown in Fig. ~\ref{fig:scal}.

\begin{figure}[hbtp]
\vspace{4. in}
\caption{Scaling of the radial cut of the structure factor}
\label{fig:scal}
\end{figure}

The anisotropy in the system neccessitates the use of an anisotropic scaling
form
where $k_c$ and $F(x)$ depend on the direction of the cut.\cite{alanis}
 With this change the
structure factor is observed to satisfy scaling during the late stages of
growth
and evolves towards Porod's law scattering where the intensity falls of with
inverse fourth power of the momentum.  The transverse cut satisfies scaling to
a
much lesser extent indicating that we are probably not in the scaling regime
for this function when we are well within the scaling regime for the radial
cuts.
As discussed earlier, this situation can arise because of the difference in
width
between the two types of domain walls.

In conclusion, we have shown that EMT provides a realistic description of
kinetics in \cutau.  The late stage growth is in qualitative agreement with
ideas
of universality and scaling.  The study of growth kinetics of first order phase
transitions extends the study of equilibrium phenomena to the study of
metastable and unstable states in an atomistic model of intermetallic alloys.

\end{document}